\documentstyle[preprint,eqsecnum,aps]{revtex}
\tightenlines
\begin{document}
%\draft %This command makes pacs numbers print
\preprint{hep-th/9709128 \hspace{9.5cm} CHIBA-EP-100 }
\title{Energy Radiation from a Moving Mirror with  Finite Mass}

\author{Riuji Mochizuki
\footnote{e-mail address:mochi@cuphd.nd.chiba-u.ac.jp},$\ $ Kenji
Ikegami\footnote{e-mail address:ikegami@cuphd.nd.chiba-u.ac.jp}} 
\address{Laboratory of Physics, Tokyo Dental College,
          Mihama-ku, Chiba 261, Japan}  
\author{and}
\author{Takayuki Suga\footnote{e-mail 
address:psuga@cuphd.nd.chiba-u.ac.jp}}
\address{Graduate School of Science and Technology,Chiba University,
 Inage-ku, Chiba 263,  Japan}
         

\date{\today
}
\maketitle
%%
%%\vskip 1truecm
%%\begin{center}
%%CHIBA-EP-70\\ hep-th/9709128\\ Septembert 1993
%%\end{center}
%%\vskip 1truecm
%%
\begin{abstract}
In this paper we study energy radiation from a moving mirror 
in 1+1 dimensional space-time.  The mirror is assumed 
to have finite mass and accordingly to receive 
back reaction from scalar photon field.  
The mode expansion of the scalar field 
becomes different from that without back reaction though the 
trajectory of the mirror is not changed.  
Then energy density of the vacuum
becomes to have finite value proportional to square 
of the mass of the mirror.  
Moreover we compute the energy momentum tensor of 
the radiation in the case that acceleration of the mirror is small. 
As a result we show that the mirror creates energy radiation 
whose quantity does not depend on its mass but on its acceleration 
even if the acceleration is uniform.   
%\break PACS number(s): 03.65.Ca, 11.10.Ef

\end{abstract}

%\pacs{\tt$\\$\string pacs\{03.65.Ca, \ 11.10.Ef\}
%Valid PACS appear here.
%{\tt$\backslash$\string pacs\{\}} 
%should always be input, 
%even if empty.
\vfill\eject

% body of paper here
\narrowtext

\section{Introduction}

The problem of radiation has entered a new and more interesting  
phase when it was known that quantum effects on vacuum may 
extract radiation from the 
vacuum.   Unruh[1] has shown that an accelerated  observer observes 
a heat bath in Minkowski vacuum, 
which relates to Hawking radiation from a black hole[2] 
and may give hints on quantum field theory in strong gravitational 
background.
 On the other hand, a star collapsing to become a black hole[2] 
 is mimicked by an exponentially accelerated mirror 
 with infinite mass[3].    
A geometric boundary made by the mirror gives vacuum
a quantum effect same as the collapsing star.  
Nevertheless these phenomena  
have right to attract attention even if the relations 
to black hole is ignored.  
For example we should consider these effects 
when studying radiation from electrons in an accelerator[4].  
Moreover  principle of equivalence should be 
reinterpreted to contain the quantum effects of radiation.  

 If we confirm a moving mirror effect experimentally, 
 we will, of course, have to use an object which has finite mass.  
In this paper we study how a moving mirror with finite mass 
(say {\it dynamical} boundary condition for scalar photons) 
affects on Minkowski vacuum in 1+1 dimensional space-time.    
We take account of back reaction on the mirror  
by considering energy and momentum conservation upon reflection 
of scalar photons[5][6].  
Mode functions of the scalar field are accordingly modified 
into forms which contain integrations by the coordinate 
and the frequency.  In the computation we do not use (mass)$^{-1}$ 
expansion to see behavior of the out-going wave 
in ultraviolet region, which is strongly 
affected by back reaction as easily guessed.  As a result
energy-momentum tensor of the out-going wave, even 
the vacuum energy part, 
becomes finite and energy radiation which is  independent of 
the mass but due to finiteness of the mass appears  whenever the
mirror is accelerated.  It is contrast to the result 
for the case that a mirror has infinite mass,
according to which result radiation does not occur 
if the acceleration of the mirror is uniform[7].

The remaining sections are organized as follows. In the second section 
we review the previous result by Fulling and Davies[7] about a moving 
mirror with infinite mass.  In order to clarify relation to the third 
section, 
boundary condition is take into account by using Doppler factor. 
Then we treat a moving mirror with finite mass in the third section.  
Energy radiation from the mirror is computed for the case that the 
acceleration of the mirror is small.  The last section is assigned 
to conclusion.

\section{A Moving Mirror with Infinite Mass}

In this section we treat a mirror with infinite mass which 
travels along a trajectory 
in two-dimensional Minkowski space-time.  The time and space 
coordinates 
$\overline t$, $\overline x$
on the trajectory of the mirror are combined as  
\begin{equation}
\overline x=z(\overline t).
\end{equation}
For simplicity we use for {\it photon}  massless scalar field 
$\phi$ which obeys 
the Klein-Gordon equation
\begin{equation}
{\partial^2 \phi\over\partial u\partial v}=0,
\end{equation}
where $u$ and $v$ are null coordinates defined as
\[
u=t-x,\ \ \ \ \ v=t+x.
\]
The scalar field is reflected by the mirror, so that a boundary 
condition 
\begin{equation}
\phi(\overline t,\overline x)=0
\end{equation}
is imposed.  Hence it is necessary that the mode expansion is 
changed from that without boundary condition.  
First we work in {\it in mode}, 
where in-coming wave traveling from the right to the
left is expanded with plane waves  and out-going wave travelling from 
the left to the right, on the other hand, may become complicated form.  
We write the mode function $\phi_{IN}$ with energy $\omega$ as
\begin{equation}
\phi_{IN}(u,v)\sim\exp[-i\omega v]-\exp[-i\omega p(u)],
\end{equation}
where
\begin{equation}
p(u)\equiv 2\overline t(u)-u.
\end{equation}
We can similarly obtain the mode function $\phi_{OUT}$ in 
{\it out mode}:
\begin{equation}
\phi_{OUT}(u,v)\sim\exp[-i\omega u]-\exp[-i\omega f(v)],
\end{equation}
where
\begin{equation}
f(v)\equiv 2\overline t(v)-v.
\end{equation}

If we regard {\it light} as classical wave, relation 
between incident and reflected 
angular frequencies
$\omega$ and
$\tilde\omega_0$ is known as (relativistic) Doppler's formula:
\begin{equation}
\tilde\omega_0 =D(V)\omega,
\end{equation}
where velocity of the mirror $V$ and Doppler factor $D$ are
\begin{equation}
V\equiv{dz(t)\over dt},
\end{equation}
\begin{equation}
D\equiv{\partial \tilde\omega_0\over\partial\omega}
={1+V\over 1-V}.
\end{equation}
The above two descriptions of the  phenomenon that light is 
reflected by a mirror, 
of course,  relates to each other.  
Exponents of the mode functions (2.5) and (2.7) 
are  expressed by the Doppler factor
$D$ as the following:
\begin{mathletters}
\begin{eqnarray}
{dp(u)\over du}&=&2{d\overline t\over du}-1\nonumber\\
&=&{2\over 1-V(u)}-1\nonumber\\
&=&D(u),
\end{eqnarray}
\begin{eqnarray}
{df(v)\over dv}&=&2{d\overline t\over dv}-1\nonumber\\
&=&{2\over 1+V(v)}-1\nonumber\\
&=&{1\over D(v)},
\end{eqnarray}
\end{mathletters}
Thus
\begin{mathletters}
\begin{equation}
\phi_{IN}\sim \exp[-i\omega v]-\exp[-i\omega\int D(u)du],
\end{equation}
\begin{equation}
\phi_{OUT}\sim\exp[-i\omega u]-\exp[-i\omega\int{dv\over D(v)}].
\end{equation}
\end{mathletters}
In this way we obtain the mode functions written by means of the 
Doppler factor, 
which functions are useful when back reaction on the mirror is 
considered.

When the mode function is different from that of plane wave, 
radiation may occur.  
We use energy-momentum tensor to investigate the radiation from 
the moving 
mirror.  The Lagrangian density of the free scalar field in the 
$t-x$ coordinate 
system is 
\begin{equation}
L={1\over 2}\partial_{\mu}\phi(t,x)\partial^{\mu}\phi(t,x).
\end{equation}
Then classical energy-momentum tensor defined as
\begin{equation}
{\rm T}^{\mu}_{\ \nu}={\partial
L\over\partial(\partial_{\mu}\phi)}\partial_{\nu}\phi
-\delta^{\mu}_{\ \nu}L,
\end{equation}
has components 
\begin{mathletters}
\begin{eqnarray}
{\rm T}_{00}&=&{\rm T}_{11}\nonumber\\
&=&{1\over 2}\big\{ (\partial_0\phi)^2+(\partial_1\phi)^2\big\}
\end{eqnarray}
\begin{eqnarray}
{\rm T}_{01}&=&{\rm T}_{10}\nonumber\\
&=&{1\over 2}\big\{ \partial_0\phi\partial_1\phi
+\partial_1\phi\partial_0\phi\big\}.
\end{eqnarray}
\end{mathletters}
After this we shall regard $\phi$ as a quantum field operator 
expanded in {\it in mode} which is defined as
\begin{equation}
\phi=\int^{\infty}_0 d\omega[a_{\omega}\phi_{\omega}
+a_{\omega}^{\dagger}\phi_{\omega}^{\ast}],
\end{equation}
where $a_{\omega}$ and $a_{\omega}^{\ast}$ are annihilation and 
creation operators, respectively
and $\phi_{\omega}$ is the mode function defined in (2.12a).  
Since in the above mode expansion the concept of {\it particle} is 
clear only 
for in-coming wave, the {\it in-vacuum} $\mid
0\rangle$ which is defined as
\begin{equation}
a_{\omega}\mid 0\rangle =0
\end{equation}
should be regarded as the vacuum with no in-coming particle.  
On the other hand, out-going particles may exist.  
If one wants to count number of the out-going 
particles, he or she can do it by, for example, computing 
Bogoliubov coefficients between in mode and out mode[8].  
But relation between the Bogoliubov coefficients 
and count of particles by a detector is not 
so clear for some cases[9][10][11].  
We will comment on this problem in the last section.  
 
Then the components of the corresponding energy-momentum tensor 
operator
$T_{\mu\nu}(t,x)$, which is obtained by substituting (2.16) 
into (2.15), are
\begin{mathletters}
\begin{equation}
\begin{array}{rll}
T_{00}=&T_{11}&\\
=&{1\over 2}\int^{\infty}_{0}
d\omega\int^{\infty}_0d\omega^{\prime}
&\big[a_{\omega}a_{\omega^{\prime}}
\partial_{\mu}\phi_{\omega}\partial_{\mu}\phi_{\omega^{\prime}}
+a_{\omega}^{\dagger}a_{\omega^{\prime}}^{\dagger} 
\partial_{\mu}\phi_{\omega}^{\ast}
\partial_{\mu}\phi_{\omega^{\prime}}^{\ast}\\
&&+2a_{\omega}^{\dagger}a_{\omega^{\prime}}
\partial_{\mu}\phi_{\omega}^{\ast}\partial_{\mu}\phi_{\omega^{\prime}}
+\delta(\omega
-\omega^{\prime})\partial_{\mu}
\phi_{\omega}^{\ast}\partial_{\mu}\phi_{\omega^{\prime}}\big]\\
\end{array}
\end{equation}
\begin{equation}
\begin{array}{rll}
T_{01}=&T_{10}&\\
=&{1\over
2}\int^{\infty}_{0}d\omega\int^{\infty}_0
d\omega^{\prime}&\big[\{a_{\omega}a_{\omega^{\prime}}
\partial_{0}\phi_{\omega}\partial_{1}\phi_{\omega^{\prime}}
+a_{\omega}^{\dagger}a_{\omega^{\prime}}^{\dagger} 
\partial_{0}\phi_{\omega}^{\ast}\partial_{1}
\phi_{\omega^{\prime}}^{\ast}\\
&&\ \ +a_{\omega}^{\dagger}a_{\omega^{\prime}}
(\partial_{0}\phi_{\omega}^{\ast}\partial_{1}\phi_{\omega^{\prime}}
+\partial_{1}\phi_{\omega}^{\ast}\partial_{0}\phi_{\omega^{\prime}})\\
&&\
\
+\delta(\omega-\omega^{\prime})\partial_{1}
\phi_{\omega}^{\ast}\partial_{0}\phi_{\omega^{\prime}}\}\\ 
&&+\{\ 0\ \leftrightarrow\ 1\ \}\big].\\
\end{array}
\end{equation}
\end{mathletters}
Expectation value of the energy-momentum tensor 
in the in-vacuum is defined as
\begin{equation}
\langle T_{\mu\nu}\rangle=T_{\mu\nu}-: T_{\mu\nu}:,
\end{equation}
and thus
\begin{mathletters}
\begin{equation}
\langle T_{00}\rangle=\langle T_{11}\rangle=
{1\over 2}\int^{\infty}_{0}
d\omega\Biggl\{(\partial_0\phi_{\omega}) 
(\partial_0\phi_{\omega}^{\ast})
+(\partial_1\phi_{\omega})
(\partial_1\phi_{\omega}^{\ast})\Biggr\},
\end{equation}
\begin{equation}
\langle T_{01}\rangle=\langle T_{10}\rangle
={1\over 2}\int^{\infty}_{0}
d\omega\Biggl\{(\partial_0\phi_{\omega}) 
(\partial_1\phi_{\omega}^{\ast})
+(\partial_1\phi_{\omega})
(\partial_0\phi_{\omega}^{\ast})\Biggr\}.
\end{equation}
\end{mathletters}
These expectation values are known to diverge quadratically for 
the mode function (2.12a).  Hence it
is necessary to regularize them[8] to obtain physical quantities, 
which problem is however out of our
scope here.  We show only the result according to Fulling and 
Davies[7]:
\begin{equation}
\langle T_{00}(u)\rangle=-\langle T_{01}(u)\rangle 
={1\over 12\pi}\Bigl\{ 
D(u)\Bigr\}^{1/2}
{d^2\{ D(u)\}^{-1/2}\over du^2},
\end{equation}
where a  constant quadratically divergent term has been discarded.  
The renormalized expectation
value (2.21) describes radiation from the mirror with no in-coming 
scalar particles and it vanishes
when acceleration of the mirror is uniform.

\section{back reaction}

If one regards a moving mirror simply as a model of a collapsing 
star, the mirror
should have infinite mass to fix its trajectory; in other words 
the boundary condition for the scalar photon field 
should be geometric.  Nevertheless, if one regards a moving 
mirror as a real object and wants to confirm 
its effect experimentally, he or she should 
take account of back reaction on the mirror.     
In this section we study how the back reaction on the 
mirror with finite mass $m$ affects energy radiation.  
We let the mirror particle and a massless scalar
field obey energy-momentum conservation law  at elastic scattering 
between them.  It is not expected
however that the scalar particle really collides with the mirror 
since we will compute physical quantities in {\it in vacuum}, 
which contains no in-coming particles.   Hence, the 
trajectory of the mirror is left unchanged though 
the mode function of the out-going 
wave is affected by the dynamical boundary condition. 

The energy and momentum conservation at the collision lead the
following equations:
\begin{mathletters}
\begin{equation}
m\gamma +\omega =m\tilde\gamma + \tilde\omega,
\end{equation}
\begin{equation}
-m\gamma V+\omega =-m\tilde\gamma\tilde V-\tilde\omega,
\end{equation}
\end{mathletters}
where $\tilde V$ and $\tilde\omega$ are velocity of the mirror and 
energy of the scalar field after the collision, respectively, and
\[
\gamma\equiv{1\over\sqrt{1-V^2}}\ ,\ \ \ \ \ \tilde\gamma
\equiv{1\over\sqrt{1-\tilde V^2}}.
\]
Regarding $\tilde\omega$ and $\tilde V$ as functions of $\omega$ and 
$V$, we obtain Doppler factor $\tilde D$ with back reaction 
\begin{eqnarray}
\tilde D&\equiv&{\partial\tilde\omega\over\partial\omega}\nonumber\\
&=&{1+\tilde V\over 1-\tilde V}\nonumber\\
&=&\Bigl\{D^{-1/2}+{2\omega\over m}\Bigr\}^{-2}.
\end{eqnarray}
If we take $m\rightarrow\infty$, $\tilde D$ becomes equal to $D$.  
For finite mass and  large
$\omega$, however,  $\tilde D$ is small and energy of the reflected 
particle is reduced severely.  It
induces convergence of energy-momentum tensor of the out-going wave, 
as we shall see later. 

Then the mode functions (2.12) will be modified  into
\begin{mathletters}
\begin{equation}
\phi_{IN}=A \exp[-i\omega v]-B\exp[-i\int\int \tilde D(u)d\omega du],
\end{equation}
\begin{equation}
\phi_{OUT}=A^{\prime}\exp[-i\tilde\omega u]-B^{\prime}
\exp[-i\int\int{d\tilde\omega dv\over
\tilde D(v)}],
\end{equation}
\end{mathletters}
where $A$, $A^{\prime}$, $B$  and $B^{\prime}$ are normalization 
factors.  

Here for convenience we introduce new notations
$\tilde\omega (\omega)$ and
$\omega(\tilde\omega )$, which are solutions of equations (3.1), 
defined as 
\begin{mathletters}
\begin{eqnarray}
\tilde\omega (\omega)&\equiv&\int \tilde D(u)d\omega\nonumber\\
&=&-{m\over 2}(D^{-1/2}+{2\omega\over m})^{-1}+{mD^{1/2}\over 2},
\end{eqnarray}
\begin{eqnarray}
\omega(\tilde\omega )&\equiv&\int{d\tilde\omega\over \tilde D(v)}
\nonumber\\
&=&+{m\over 2}(D^{1/2}-{2\tilde\omega\over m})^{-1}-
{m D^{-1/2}\over 2}.
\end{eqnarray}
\end{mathletters}
Thus
\begin{mathletters}
\begin{equation}
\phi_{IN}=A\exp[-i\omega v]-B\exp[-i\int\tilde\omega (\omega) du],
\end{equation}
\begin{equation}
\phi_{OUT}=A^{\prime}\exp[-i\tilde\omega u]-
B^{\prime}\exp[-i\int\omega(\tilde\omega )dv].
\end{equation}
\end{mathletters}
These modified forms of the mode functions are natural 
extension of (2.12) because the out-going
(in-coming) wave of  in (out) mode, as expected, 
becomes plane wave with the appropriate energy and
 momentum  if velocity of the mirror is constant.  

  To discuss orthonormality condition of the functions (3.3a), 
let us compute the Klein-Gordon inner
product of the out-going wave without the normalization factor $B$.  
\begin{eqnarray}
(&&\exp[i\int\tilde\omega (\omega_1)du],
\exp[-i\int\tilde\omega (\omega_2)du])\nonumber\\
=&&i\int
dx\Bigl\{\exp[i\int\tilde\omega (\omega_1)du]
\overrightarrow{\partial_t}
\exp[-i\int\tilde\omega (\omega_2)du]
-\exp[i\int\tilde\omega (\omega_1)du]\overleftarrow{\partial_t}
\exp[-i\int\tilde\omega (\omega_2)du]\Bigr\}\nonumber\\
=&&\int dx(\tilde\omega (\omega_1) +\tilde\omega (\omega_2))\exp[i\int
(\tilde\omega (\omega_1)-\tilde\omega (\omega_2))]du.
\end{eqnarray}
If we recall the definition of $\tilde\omega (\omega)$, 
we obtain two convenient equations:
\[
\tilde\omega (\omega_1)-\tilde\omega (\omega_2)
=(\omega_1-\omega_2)\tilde D(\omega_1,\omega_2),
\]
\[
\tilde\omega (\omega_1)+\tilde\omega (\omega_2)
=\tilde D(\omega_1,\omega_2)\Biggl\{ (\omega_1
+\omega_2)+{4D^{1/2}\omega_1\omega_2\over m}\Biggr\},
\]
where
\begin{equation}
\tilde D(\omega_1,\omega_2)\equiv \Big(D^{-1/2}+{2\omega_1\over
m}\Big)^{-1}\Big(D^{-1/2}+{2\omega_2\over m}\Big)^{-1}.
\end{equation}
These equations enable us to proceed with the computation as
\begin{eqnarray}
(3.6)&=&\int dx\tilde D(\omega_1,\omega_2)\Biggl\{ (\omega_1
+\omega_2)+{4D^{1/2}\omega_1\omega_2\over m}\Biggr\}
\exp [i(\omega_1-\omega_2)\int
\tilde D(\omega_1,\omega_2)]du\nonumber\\
&=&(\omega_1+\omega_2)\int dX\exp [i(\omega_1-\omega_2)X]+
{4\omega_1\omega_2\over m}\int dXD^{1/2}
\exp [i(\omega_1-\omega_2)X]\nonumber\\
&=&4\pi\omega_1\delta(\omega_1-\omega_2)+
{4\omega_1\omega_2\over m}\int dXD^{1/2}
\exp [i(\omega_1-\omega_2)X],
\end{eqnarray}
where
\[
X\equiv\int\tilde D(\omega_1,\omega_2)du.
\]
The last term in the last line of (3.8), which is due to 
the finiteness of the mass of the mirror,
does not produce Dirac's delta function unless
$D$ is constant (no acceleration).  If $D$ is constant, 
the mode expansion is exactly determined.  In
this case the out-going wave becomes plane wave whose frequency 
has upper bound $mD^{1/2}/2$ and the
energy-momentum tensor of the out-going wave has finite value 
$Dm^2/32\pi$.  We think it
appropriate to regard divergence which appears when the mass goes 
to infinity as discarded
constant divergence mentioned in the previous section.    
Hence the last term in the line of (3.8)
gives energy of the vacuum and should be discarded when one computes 
physical quantities.     
 
In general it is
necessary to carry out the integration and let the function satisfy 
orthonormality condition  in
order to determine the mode function for each trajectory of a mirror.  
Here we study the case that
acceleration of the mirror is small, so that the mode function in 
in mode is approximately given by (3.3a) with {\it function} $B$:
\begin{equation}
B^{-2}(u)=2\pi\Bigl\{2\omega+{4\omega^2D^{1/2}(u)\over m}\Bigr\}.
\end{equation}  
We may expect that the modification on the mode expansions due to 
the back reaction
will cause radiation of the scalar particles.  
Computing energy-momentum tensor of the
out-going wave, we show what effects the back reaction give 
to the radiation.  The mode function
of the out-going wave in in mode is written as 
\begin{eqnarray}
\phi_{IN}^{out-going}&\equiv&\varphi\nonumber\\
&=&{-1\over\sqrt{4\pi\omega}}\Biggl(1+{2\omega D^{1/2}(u)\over
m}\Biggr)^{-1/2}
\exp[-i\int\tilde\omega (\omega)du].
\end{eqnarray}
Then
\begin{mathletters}
\begin{equation}
\left.
\begin{array}{r}
\partial_0\varphi\\
\partial_1\varphi\\
\end{array}\right\}
={\pm 1\over\sqrt{4\pi\omega}}\Biggl(1+{2\omega D^{1/2}\over
m}\Biggr)^{-3/2}\Biggl\{{\omega D^{-1/2}\dot{D}\over 2m}
+i\Big(1+{2\omega D^{1/2}\over
m}\Big)\tilde\omega \Biggr\}\exp[-i\int\tilde\omega du],
\end{equation}
\begin{equation}
\left.
\begin{array}{r}
\partial_0\varphi^{\ast}\\
\partial_1\varphi^{\ast}\\
\end{array}\right\}
={\pm 1\over\sqrt{4\pi\omega}}\Biggl(1+{2\omega D^{1/2}\over
m}\Biggr)^{-3/2}\Biggl\{{\omega D^{-1/2}\dot{D}\over 2m}
-i\Big(1+{2\omega D^{1/2}\over
m}\Big)\tilde\omega \Biggr\}\exp[+i\int\tilde\omega du],
\end{equation}
\end{mathletters}
where the dot means differentiation by $u$.
Substituting them into (2.20), we obtain the energy-momentum tensor 
of the out-going wave:
\begin{eqnarray}
\langle T_{00}\rangle&=&\langle T_{11}\rangle=-\langle T_{01}\rangle
=-\langle
T_{10}\rangle\nonumber\\
&=&\int^{\infty}_0 d\omega{1\over 4\pi\omega}
\Biggl(1+{2\omega D^{1/2}\over
m}\Biggr)^{-3}\Biggl({\omega^2\dot{D}^2
\over 4m^2 D}+\omega^2 D^2\Biggr)\nonumber\\
&=&{1\over 16\pi}\Biggl({Dm^2\over 2}
+{\dot{D}^2\over 8D^2}\Biggr)\nonumber\\
&=&{Dm^2\over 32\pi}+{\gamma^2 \alpha^2\over 32\pi},
\end{eqnarray}
where
\[
\alpha\equiv \gamma\dot V.
\]
In this calculation no regularization is needed since 
the high frequency part is strongly suppressed
by recoil of the mirror.  The first term in the last line of (3.12) 
is, as previously mentioned, 
energy-momentum tensor for out-going plane wave.    
The second term  does not appear  if we put the
mass of the mirror infinite from the beginning of the computation, 
though it is independent of the
mass.  We regard this term as energy radiation from the mirror 
when no in-coming photon
exist.    

\section{Conclusion}

As well known, boundaries may extract radiation from vacuum.  
In this paper we have treated the
boundary made of a particle with finite mass (dynamical boundary), 
which particle obeys energy and
momentum conservation law when it collides with a virtual scalar 
photon.  Mode expansion of the
photon is accordingly changed to have upper bound of its energy.  
Moreover the dynamical boundary
causes energy radiation independent of the mass and proportional 
to square of acceleration of the
mirror.   Such effects are not perturbative compensation 
for geometrical boundary condition but 
novel aspects of the dynamical one.  Thus in some cases it may be 
appropriate and essential to impose dynamical 
boundary condition instead of geometric one.  

When a detector is
introduced to count number of particles,  
we should similarly remember that the detector has
finite mass.  For example a rotating detector in Minkowski vacuum 
may response though the
rotating vacuum is  the same.  This is not a quantum effect 
but a rather trivial one:  Radiation are
caused using energy which the detector has lost with recoil.  
It does not vanish even if mass of the
detector is thought to be infinity at the end of the computation[12].

%{\bf ACKNOWLEDGMENTS}

% now the references. delete or change fake bibitem. delete next three
%   lines and directly read in your .bbl file if you use bibtex.
%
%

% \begin{figure}
% \caption{}
% \label{}
% \end{figure}
%
%

% \begin{table}
% \caption{}
% \label{}
% \end{tabular}
% \end{table}

\end{document}